%% file: eprint.tex
\pdfoutput=1
\documentclass[10pt]{article}
\usepackage{graphicx}

\def\Title#1{\begin{center} {\Large #1 } \end{center}}
\def\Author#1{\begin{center}{ \sc #1} \end{center}}
\def\Address#1{\begin{center}{ \it #1} \end{center}}

\newcommand\pubblock{\rightline{\begin{tabular}{l} Proceedings of the Fifth Annual LHCP\\ \pubnumber\\
         \pubdate  \end{tabular}}}

\newenvironment{Abstract}{\begin{quotation} \begin{center} 
             \large ABSTRACT \end{center}\bigskip 
      \begin{center}\begin{large}}{\end{large}\end{center} \end{quotation}}

\newenvironment{Presented}{\begin{quotation} \begin{center} 
             PRESENTED AT\end{center}\bigskip 
      \begin{center}\begin{large}}{\end{large}\end{center} \end{quotation}}

\input econfmacros.tex

\usepackage{color}

\newcommand\pubnumber{ CMS CR-2017/208 }

\newcommand\pubdate{\today}

\def\affiliation{
On behalf of the ATLAS and CMS Collaborations, \\
Institute of High Energy Physics, \\
Austrian Academy of Sciences, 1050 Vienna, Austria}

\providecommand*\USP{~}
\newcommand {\tauhad}    {\ensuremath{\tau_\mathrm{had}}}
\newcommand*{\UGeV}{\ensuremath{\USP\mathrm{GeV}}}
\def\pt{p_\mathrm{T}}
\def\ptt{\pt^{\tau}}

\begin{document}

\large
\begin{titlepage}
\pubblock

\vfill
\Title{  Identification and energy calibration of hadronic $\tau$ lepton decays at the LHC  }
\vfill

\Author{ Martin Flechl  }
\Address{\affiliation}
\vfill
\begin{Abstract}
The identification of hadronic $\tau$ lepton decays is an important requirement for the LHC physics program, both in 
terms of standard model measurements and the search for beyond-the-standard-model physics. The ATLAS and CMS algorithms for identification 
and energy calibration of hadronic $\tau$ lepton decays are conceptually different. They are described together with 
measurements of relevant performance figures.
\end{Abstract}
\vfill

\begin{Presented}
The Fifth Annual Conference\\
 on Large Hadron Collider Physics \\
Shanghai Jiao Tong University, Shanghai, China\\ 
May 15-20, 2017
\end{Presented}
\vfill
\end{titlepage}
\def\thefootnote{\fnsymbol{footnote}}
\setcounter{footnote}{0}
%

\normalsize 


\section{Introduction}
A robust and well-performing identification of hadronic $\tau$ lepton decays (\tauhad) is an essential ingredient 
to many searches and measurements at ATLAS~\cite{atlas} and CMS~\cite{cms}. Most prominently, Higgs boson decays to $\tau$ leptons 
provide the highest direct sensitivity for Yukawa couplings in the standard model (SM). Reconstructing these decays allows to 
measure the coupling strength of Higgs bosons to $\tau$ leptons as well as its tensor structure (CP properties) and 
off-diagonal Yukawa coupling matrix elements (e.g. in lepton-flavor-violating $H \to \tau\mu$ decays). 
Decays to $\tau$ leptons are the most sensitive probe for additional Higgs bosons of the minimal supersymmetric 
extension of the SM in most of the parameter space, and $\tau$ leptons also allow to explore 
extended gauge groups via decays of additional heavy gauge bosons as well as leptoquark decays. In order to 
optimize the sensitivity of these analyses, the two most important properties of hadronic $\tau$ lepton decay reconstruction 
are a high efficiency of reconstructing these decays (while keeping misidentification at an acceptable level) 
and an energy calibration close to the true scale with decent momentum resolution. Dedicated algorithms of the ATLAS and CMS 
experiments and their performance are explored in the following.

\section{Tau decay properties}
The mean life time of a $\tau$ lepton is $2.9\cdot10^{-13}$ s which means 
that e.g. with an energy of $50 \UGeV$, 
it travels on average 3~mm before decaying.
Most of the time, it decays before reaching the innermost sensitive detector layer and only its decay products are observed.
The $\tau$ lepton mass is $1.777 \UGeV$. It decays to hadrons plus a neutrino in 65\% of all cases. 
Among the hadronic decays, 18\% are to one charged hadron, 57\% to one charged hadron and additional neutral hadrons, 
and 23\% to three charged hadrons and any number of neutral hadrons.

\section{Tau Identification}
The detector signature of a hadronic $\tau$ lepton decay are one or three tracks
and energy depositions in the electromagnetic and (to a lesser extent) hadronic calorimeters.
Due to their large cross section, the main background for $\tau$ identification
in hadron colliders are hadronic jets, initiated by quarks and gluons.
Electrons are also frequently misidentified as \tauhad.

Hadronic $\tau$ lepton decays have a lower track multiplicity than hadronic jets,
and they are more collimated.
Compared to hadronic jets, \tauhad\ have a larger
electromagnetic component since their $\pi^0$ content is on average higher.
This is particularly important for 1-prong $\tau$ decays where the branching ratio to
$\pi^0$ states is large. Furthermore, unlike gluons and light quarks, $\tau$ leptons
do not decay directly at the interaction point and thus an impact parameter incompatible 
with zero within resolution and,
in case of three-prong decays, a secondary vertex can be reconstructed.

\subsection{ATLAS TauID}
\begin{figure}[htb]
\centering
\includegraphics[width=0.40\textwidth]{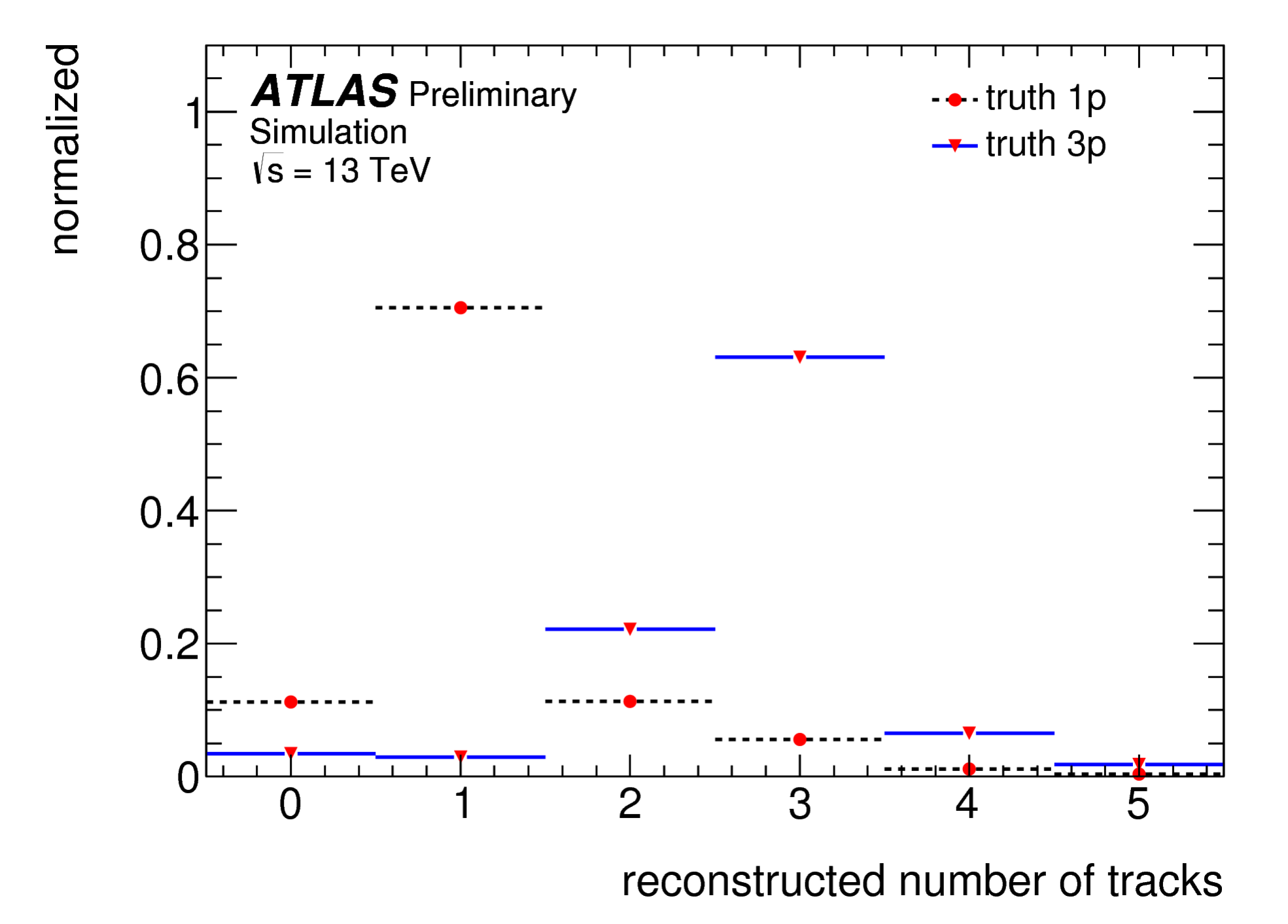} 
\includegraphics[width=0.29\textwidth]{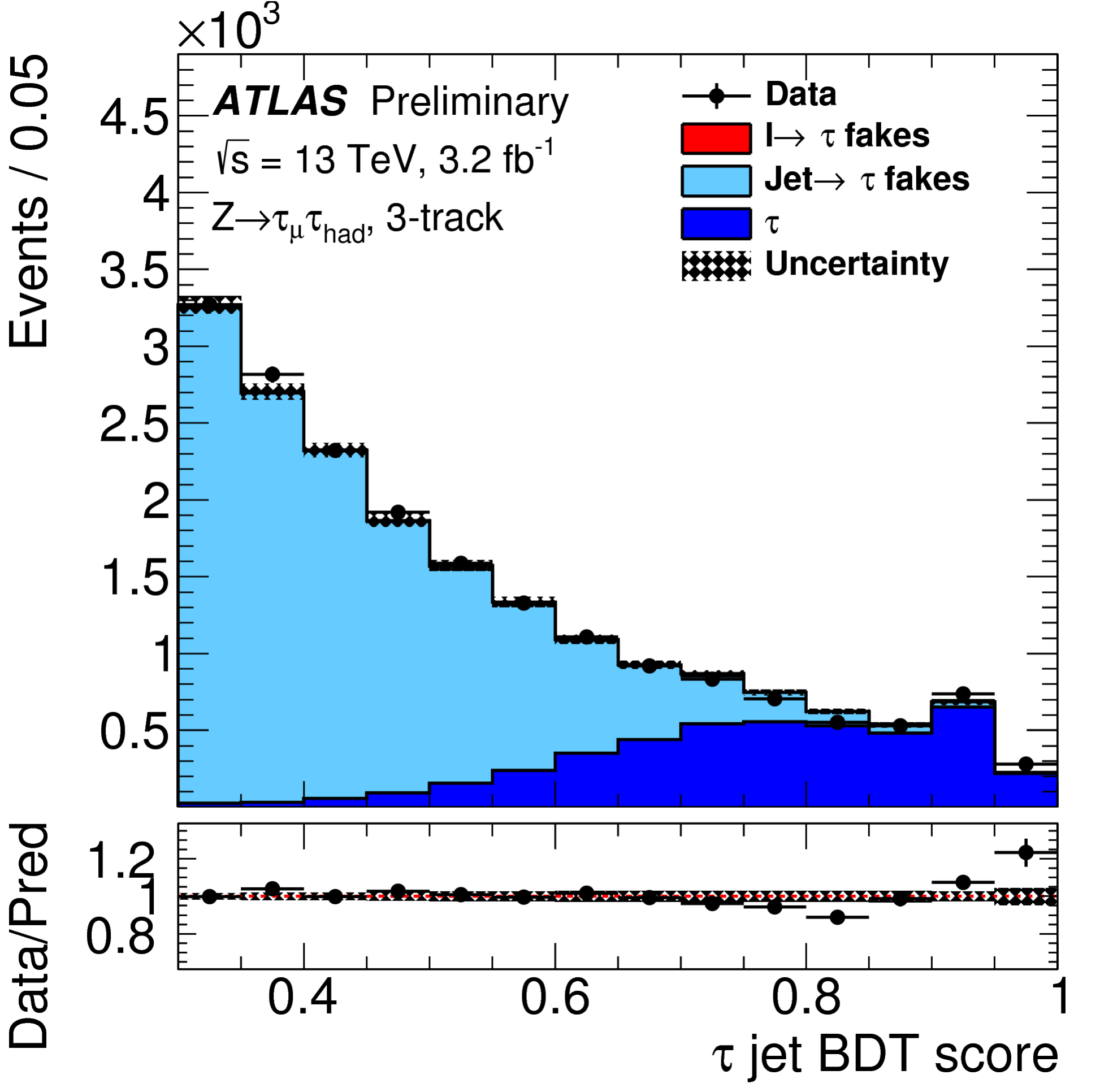} 
\includegraphics[width=0.29\textwidth]{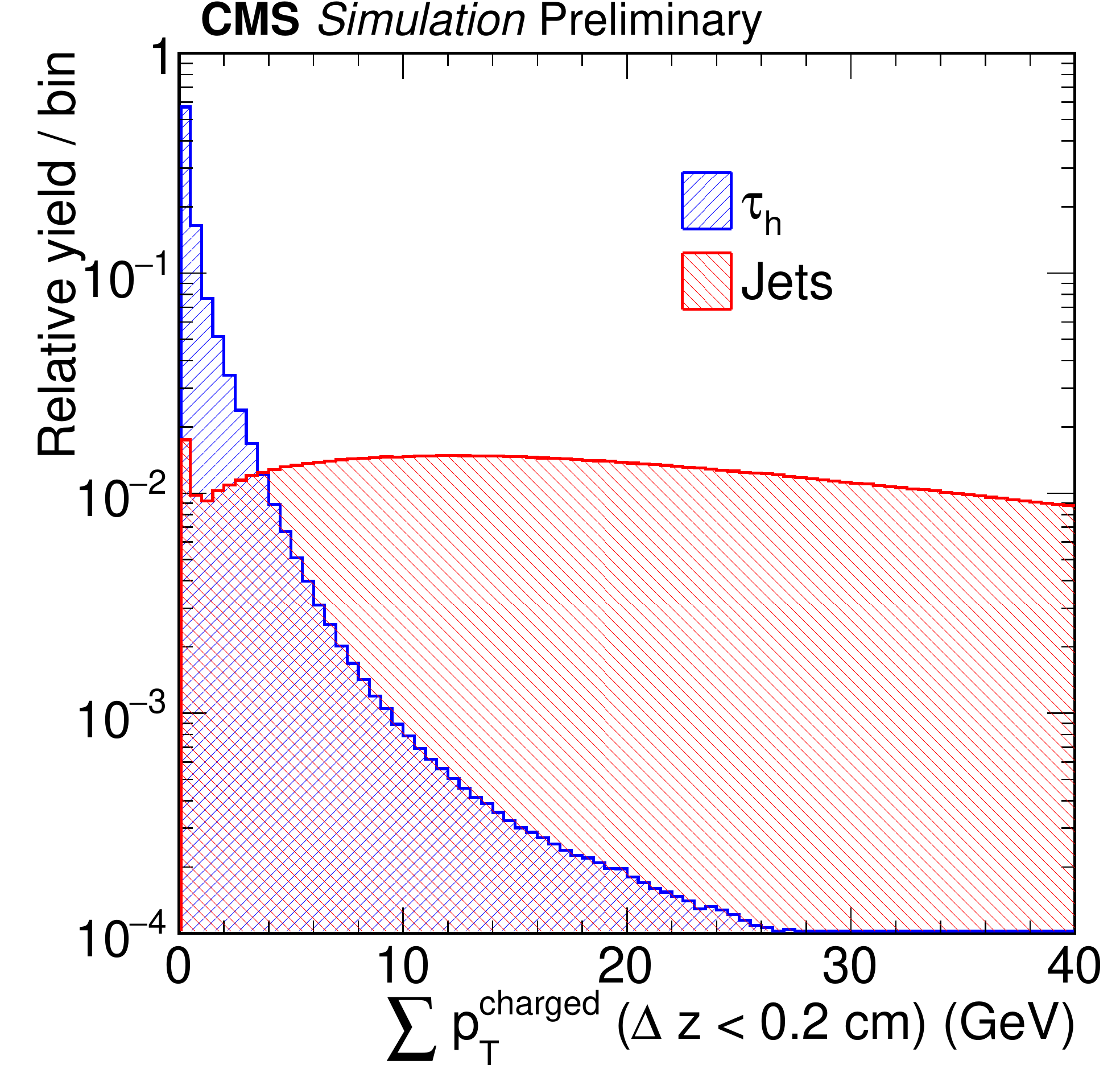} 
\caption{
Reconstructed number of tracks depending on the true number of charged hadrons for ATLAS~\cite{atlas_run2} (left). 
ATLAS BDT score distribution for 3-track \tauhad\, after requiring 
a selection to enhance $Z \to \tau\tau \to \mu \tauhad+3\nu$~\cite{ATLAS:2017mpa} (center).
CMS charged isolation sum~\cite{CMS:2016gvn} (right).
}
\label{fig:atlas_id1}
\end{figure}
The ATLAS tau algorithm~\cite{atlas_run2} consists of two steps. First, \tauhad\ candidates are reconstructed.
Then, during the TauID step, \tauhad\ are discriminated from background objects. 
The $\tau$ reconstruction uses jets with $\pt>10 \UGeV$ and $|\eta|<2.5$ as input.
These jets are formed using the anti-$k_t$ algorithm, with a
distance parameter value of 0.4 applied to clusters of calorimeter cells,
and calibrated using a local hadronic calibration.
The candidate momentum is built using only clusters of calorimeter cells within $\Delta R=0.2$ of the jet seed direction.
Tracks are associated to the candidate if they are within $\Delta R=0.2$ of the candidate direction,
have $\pt>1 \UGeV$ and fulfill certain track quality requirements. The efficiency
to reconstruct \tauhad\ with the correct number of tracks is roughly 70\% for $\ptt < 200 \UGeV$, 
see Figure~\ref{fig:atlas_id1}, and provides very little rejection against hadronic jets and light leptons.

The TauID is based on the machine-learning algorithm boosted decision tree (BDT),
using input variables calculated directly from reconstructed tracks and calorimeter depositions.
BDTs are trained separately for one-track and three-track candidates. 
In total 12 input variables are used. They are related to
how collimated the tracks or calorimeter energy depositions are, the impact parameter or secondary vertex
associated to the $\tau$ lepton decay, the invariant mass of a subset of the $\tau$ lepton decay products,
the fraction of energy deposited in the ECAL, and the balance of track momentum and calorimeter energy measurements.
The input variables are corrected for the average effect of concurrent proton-proton interactions (pile-up).
A resulting BDT score distribution is shown in Figure~\ref{fig:atlas_id1}.
%

\subsection{CMS TauID}
The CMS TauID~\cite{CMS:2016gvn} is conceptually different from the ATLAS version as it does not start with relatively low-level detector 
information (tracks, calorimeter depositions) but rather with objects reconstructed by a particle flow (PF) algorithm, 
namely electrons, muons, photons, and charged and neutral hadrons.
Unlike for ATLAS TauID, individual hadronic decay modes are reconstructed by combining PF objects.

The reconstruction step is seeded by jets reconstructed by the anti-$k_t$ algorithm with a distance parameter of 0.4 with $\pt>14 \UGeV$ and $|\eta|<2.5$.
Then, combinations of charged and neutral PF objects within the jet compatible with \tauhad\ decays are identified.
The high probability of photons from $\pi^0 \to \gamma\gamma$ decays to convert to $e^+ e^-$ pairs is taken into account by
clustering photon and electron constituents. Since 2015, a dynamic strip size depending on the $\pt$ of the candidates is used. 
Strips with a $\pt$ sum of electrons
and photons above $2.5 \UGeV$ are considered as $\pi^0$ candidates.

All combinations of the six charged particles and six $\pi^0$ candidates leading in $\pt$ are considered if the sum has unity electric charge
and all constituents are within a cone of $\Delta R = 3 \UGeV / \pt$ of the candidate direction,
with a minimum of 0.05 and a maximum of 0.1.
Each combination has to be compatible within a mass window with a resonance responsible for the particular decay mode.
All major decay modes are considered, except for 3-track decays with additional neutral hadrons.

The main handle to reduce the large background from hadronic jets are isolation requirements. Two algorithms are employed: A cut-based isolation; and
MVA isolation which uses other variables in addition to isolation quantities. The isolation sum, illustrated in Fig~\ref{fig:atlas_id1}, is defined as
\begin{equation}\label{eq:iso}
\sum \pt^{\mathrm{charged}}(d_Z<0.2\, \mathrm{cm}, \Delta R<0.5) + \mathrm{max} \left\{ 0 ,  \sum \pt^\gamma(\Delta R<0.3) - \Delta\beta \sum \pt^{\mathrm{charged}}(d_Z>0.2\, \mathrm{cm}, \Delta R<0.8) \right\},
\end{equation}
where $d_Z$ is the longitudinal distance to the \tauhad\ production vertex, 
and $\Delta\beta$ is 0.2.
Particles used in the \tauhad\ reconstruction are not considered.
The last term is an estimator of the pile-up contribution to the neutral isolation component, based on its charged component.

\begin{figure}[htb]
\centering
\includegraphics[width=0.32\textwidth]{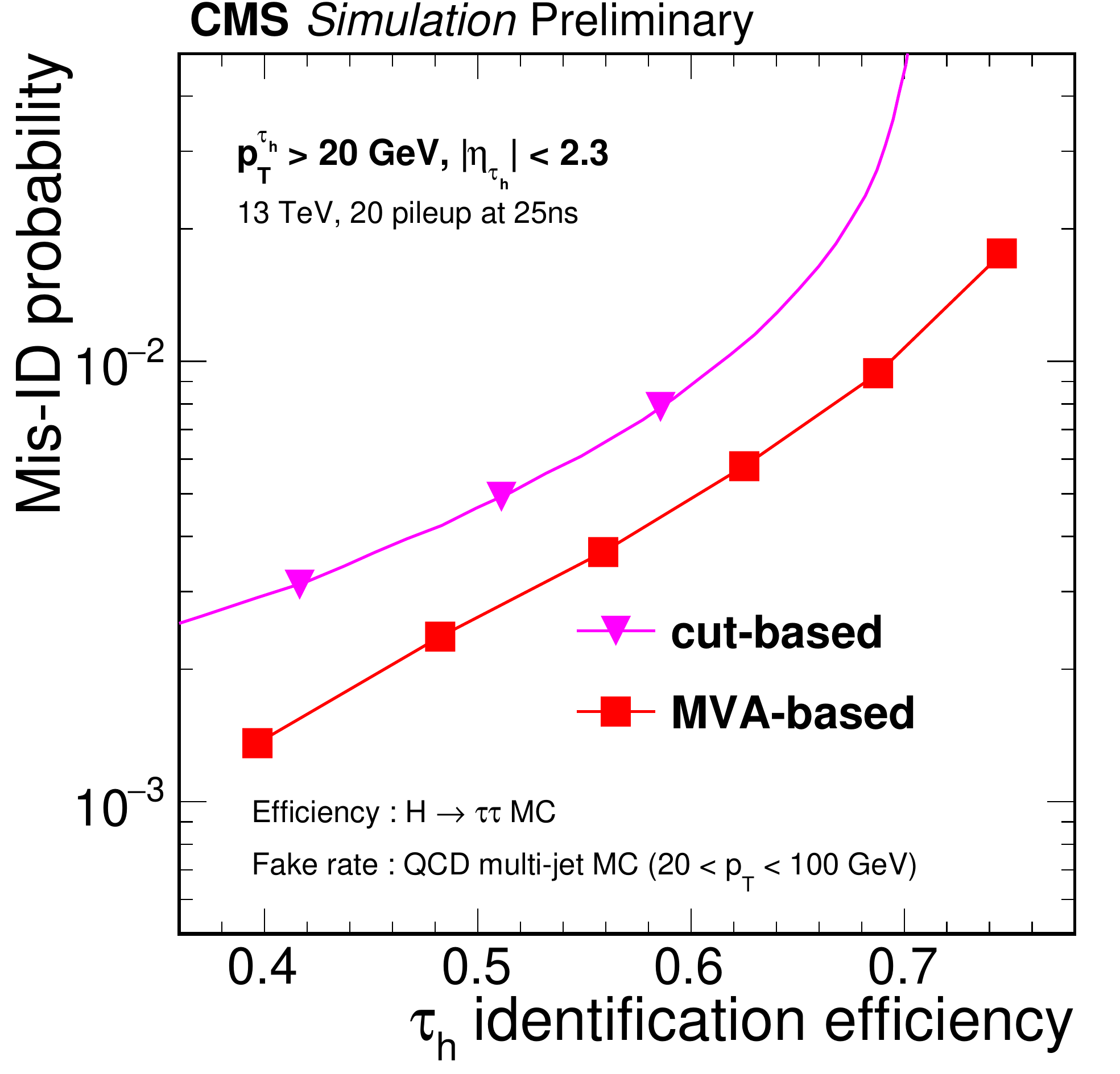}
\includegraphics[width=0.33\textwidth]{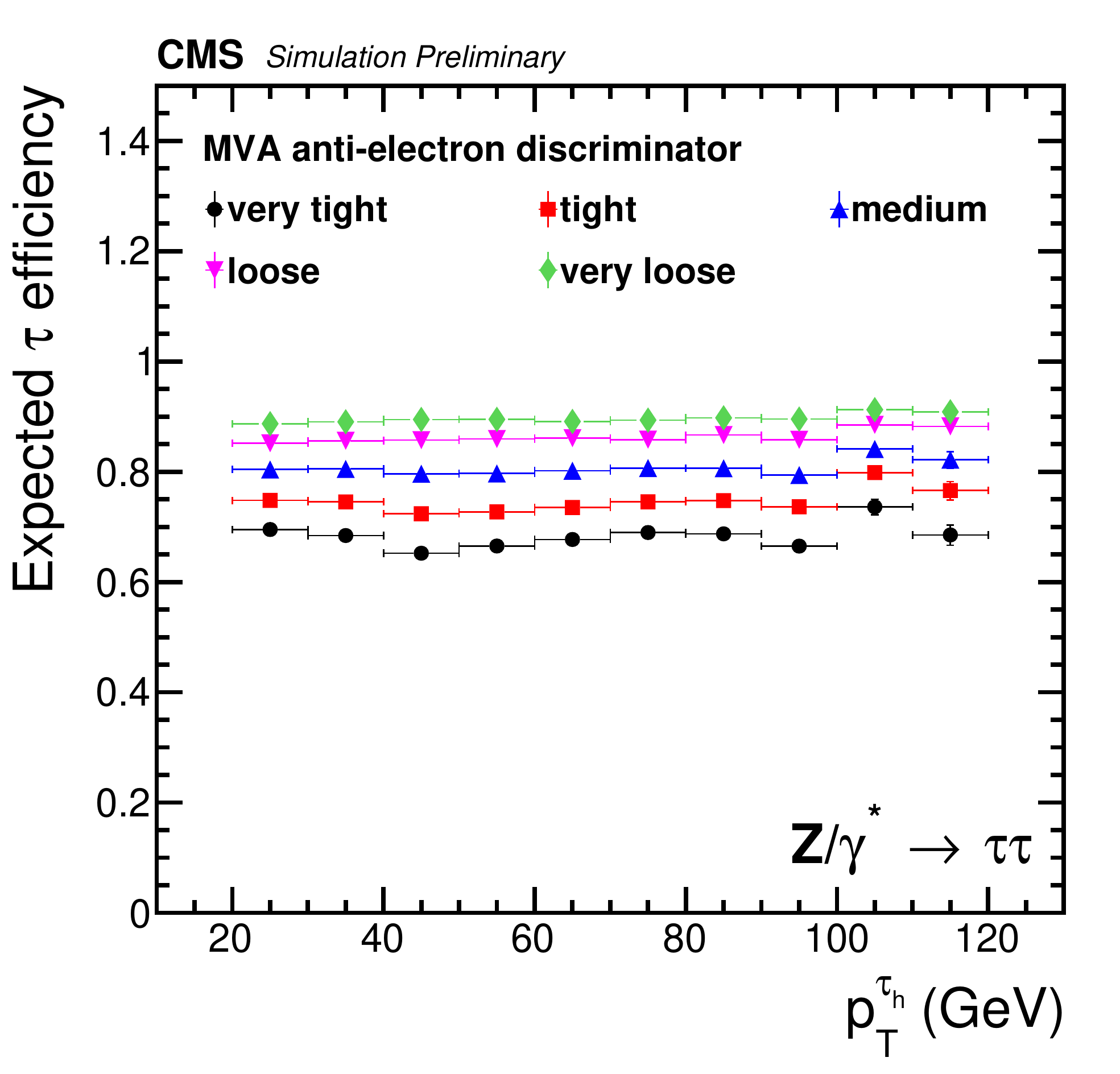}
\includegraphics[width=0.33\textwidth]{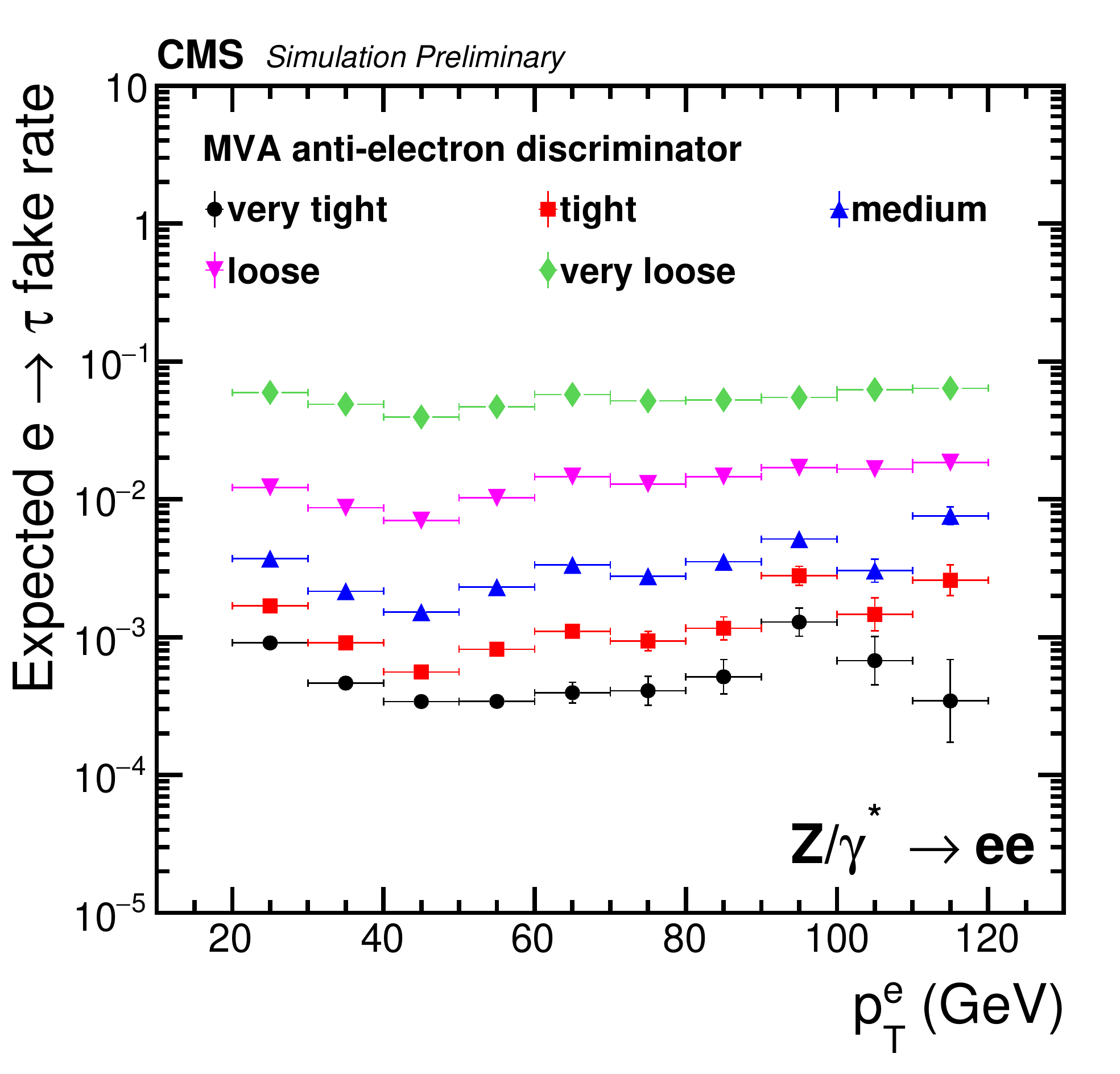}
\caption{Expected performance of CMS isolation discriminators (left). 
CMS Electron suppression: \tauhad\ efficiency (center) and electron misidentification rate (right)~\cite{CMS:2016gvn}.
}
\label{fig:cms_id1}
\end{figure}
The MVA isolation uses a BDT with the following input variables: separately, the isolation terms defined in equation~(\ref{eq:iso}); the reconstructed $\tau$ decay
mode; the signed impact parameter of its highest-$\pt$ track as well as the ratio to its uncertainty; and variables related to shape and multiplicity 
of the photon and electron content in signal and isolation cones. For three-track candidates, in addition information related
to the distance between primary and \tauhad\ decay vertex is used. The BDT is trained on simulated samples. 
The expected performance of the isolation discriminators is shown in Fig.~\ref{fig:cms_id1}.

\subsection{Light-lepton suppression}
Electrons have characteristics similar to 1-prong $\tau$ decays: one charged
track, and significant energy deposition in the calorimeter. In ATLAS, this background is suppressed by inverting the regular electron identification. 
CMS uses a BDT-based discriminator. Typical working points are about 80\% efficient for electron misidentification rate of $10^{-2}-10^{-3}$, see Figure~\ref{fig:cms_id1}.
Muon misidentification is less frequent and typical discriminators have an efficiency of $95\%-100\%$. CMS optionally vetoes \tauhad\ candidates if matching 
segments exist in the muon detector, and ATLAS if they geometrically overlap with identified muons.


\section{Tau energy calibration}
\subsection{ATLAS calibration}
\begin{figure}[htb]
\centering
\includegraphics[width=0.39\textwidth]{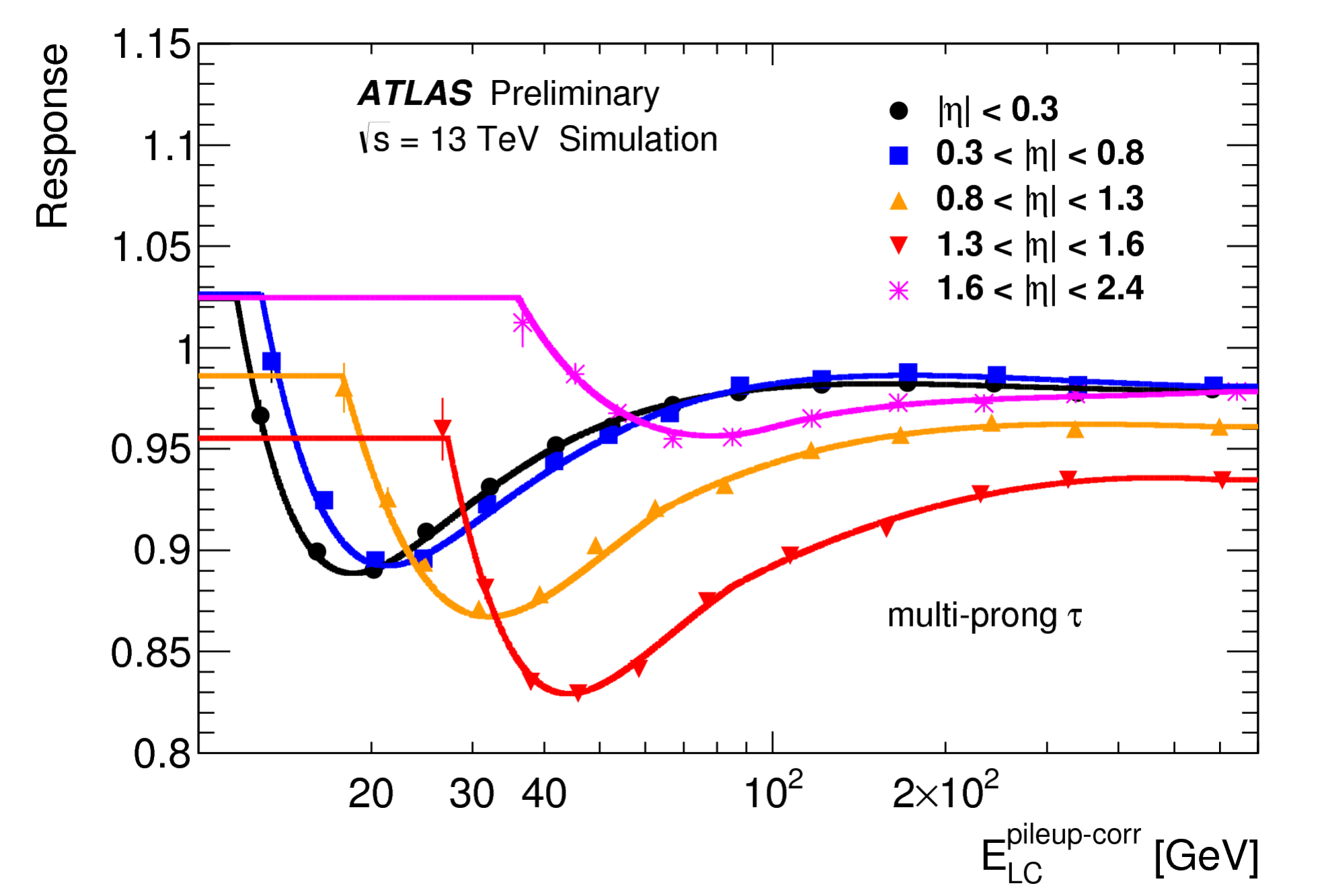} 
\includegraphics[width=0.32\textwidth]{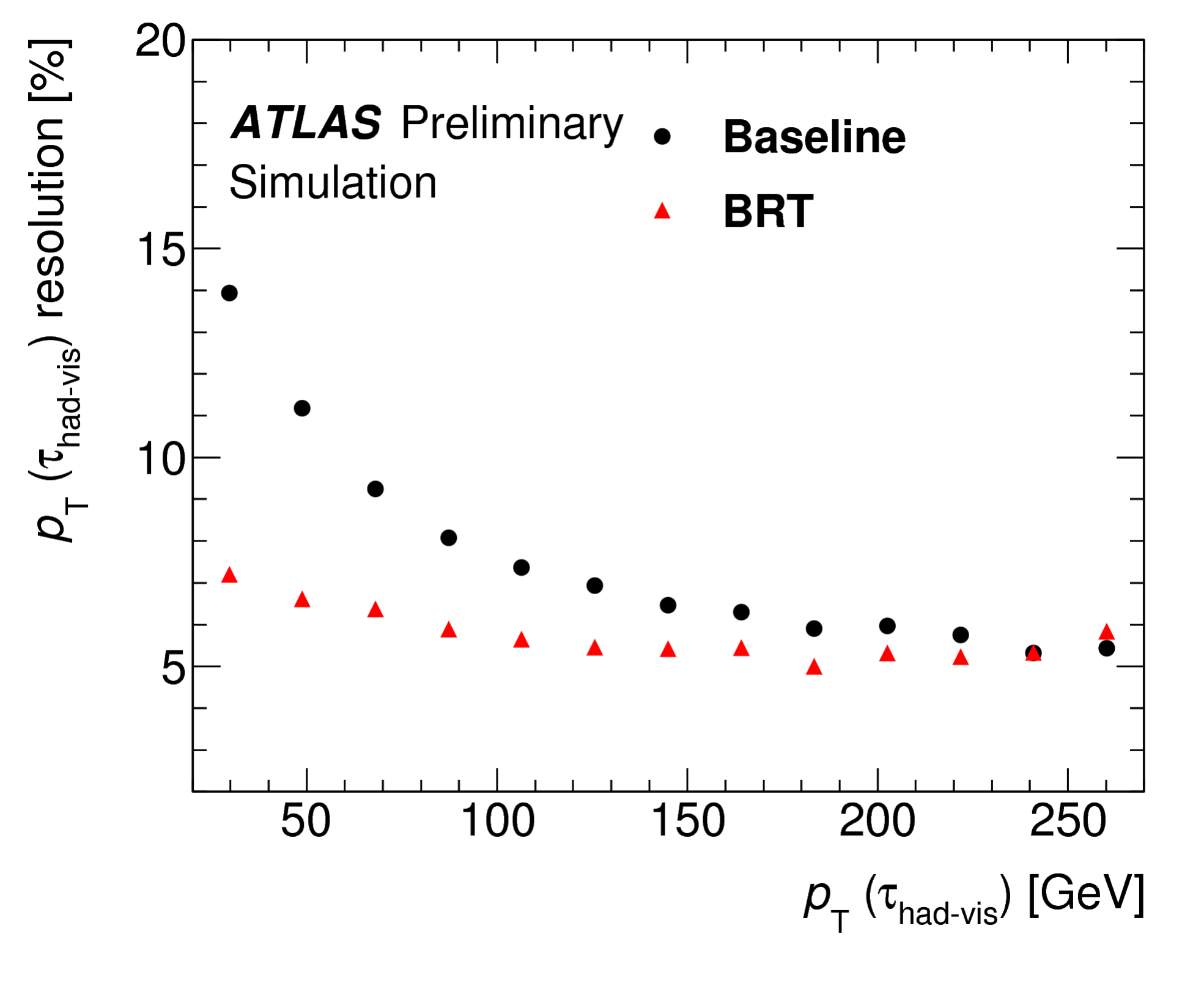} 
\includegraphics[width=0.27\textwidth]{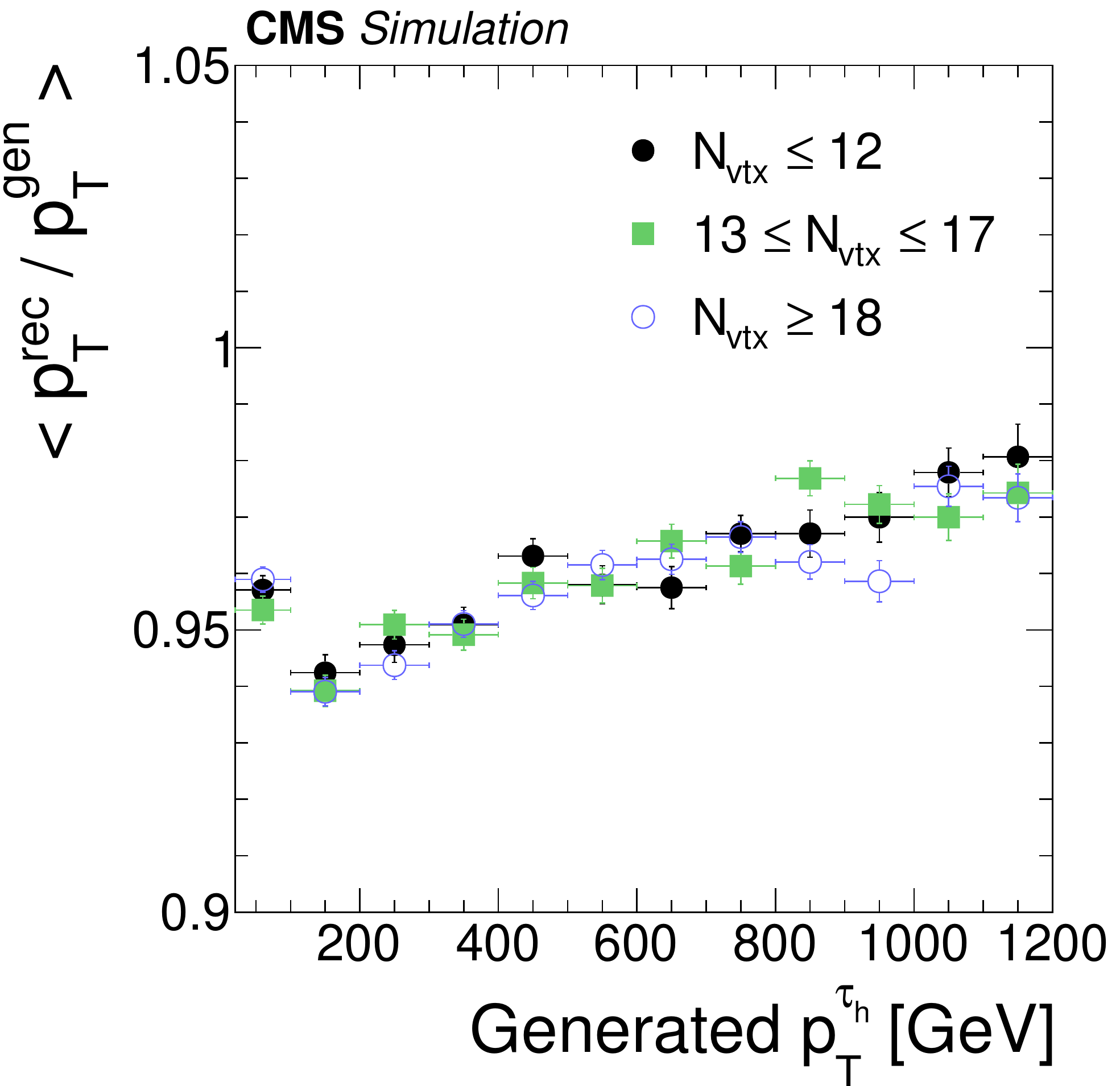} 
\caption{
ATLAS correction function used to calibrate the $\tauhad$ momentum for 3-prong decays~\cite{atlas_run2} (left). 
ATLAS \tauhad\ momentum resolution~\cite{ATLAS:2017mpa} (center), comparing the calorimeter-based approach to a boosted regression tree (BRT) using in addition 
hadronic $\tau$ decay products as input. CMS \tauhad\ response~\cite{Khachatryan:2015dfa} (right).}
\label{fig:atlas_cms_cal}
\end{figure}

The \tauhad\ candidates use jets employing a local hadronic calibration as input. This primarily corrects for the non-compensating
nature of the calorimeter and for energy depositions outside the reconstructed clusters and in non-instrumented regions of the detector.
However, it is not optimized for the cone size used to measure the $\tauhad$ momentum ($R=0.2$) and the specific mix of hadrons
in $\tau$ lepton decays, nor does it correct for effects of the underlying event and pile-up. For this reason, 
a correction function~\cite{atlas_run2} (see Figure~\ref{fig:atlas_cms_cal}) is derived using simulated samples in which the offset between reconstructed and true energy is
removed (on average) in bins of $E$, $\eta$ and number of tracks. Additionally, pile-up corrections are applied. 

Recently, an alternative energy calibration algorithm has been implemented, based on reconstructing the hadronic $\tau$ decay
products~\cite{Aad:2015unr}. This significantly improves the expected $\tau$ energy resolution at low $\pt$, e.g. from $14\%$ to
$7\%$ at $\pt=30 \UGeV$. For $\pt>200 \UGeV$, both methods yield a comparable resolution of about $5\%$, see Figure~\ref{fig:atlas_cms_cal}.

\subsection{CMS calibration}
\begin{figure}[htb]
\centering
\includegraphics[width=0.29\textwidth]{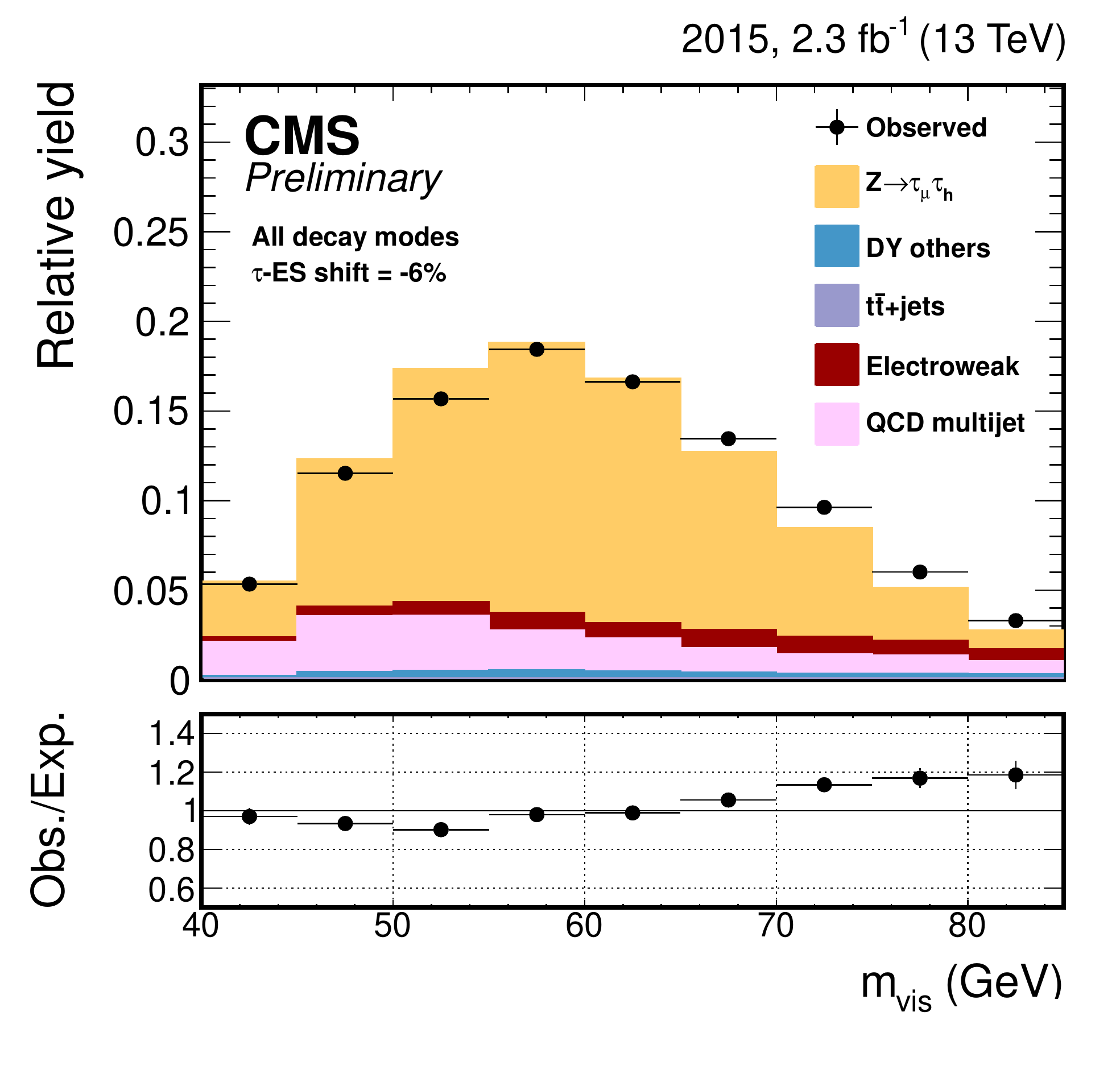}
\includegraphics[width=0.29\textwidth]{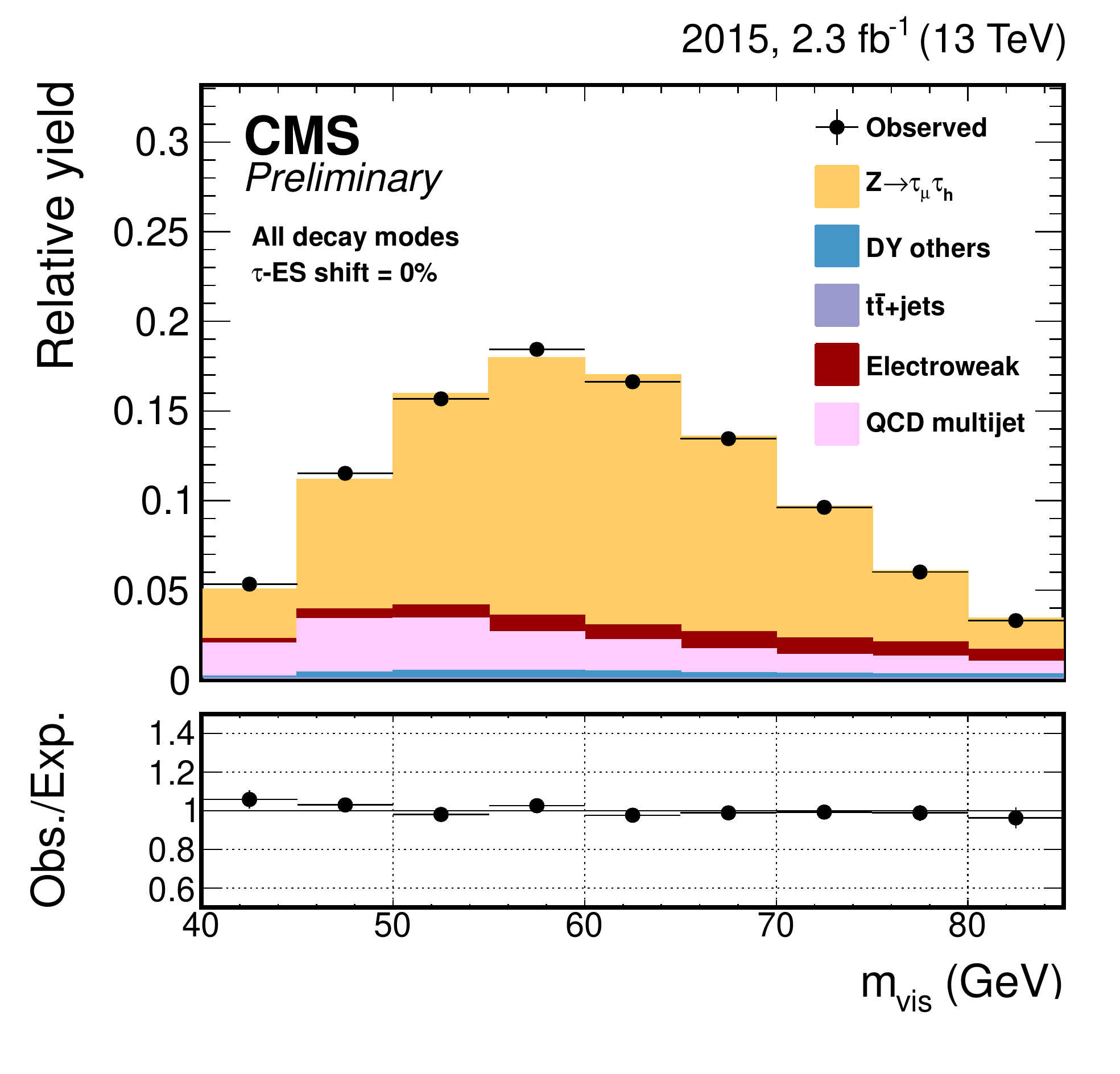}
\includegraphics[width=0.29\textwidth]{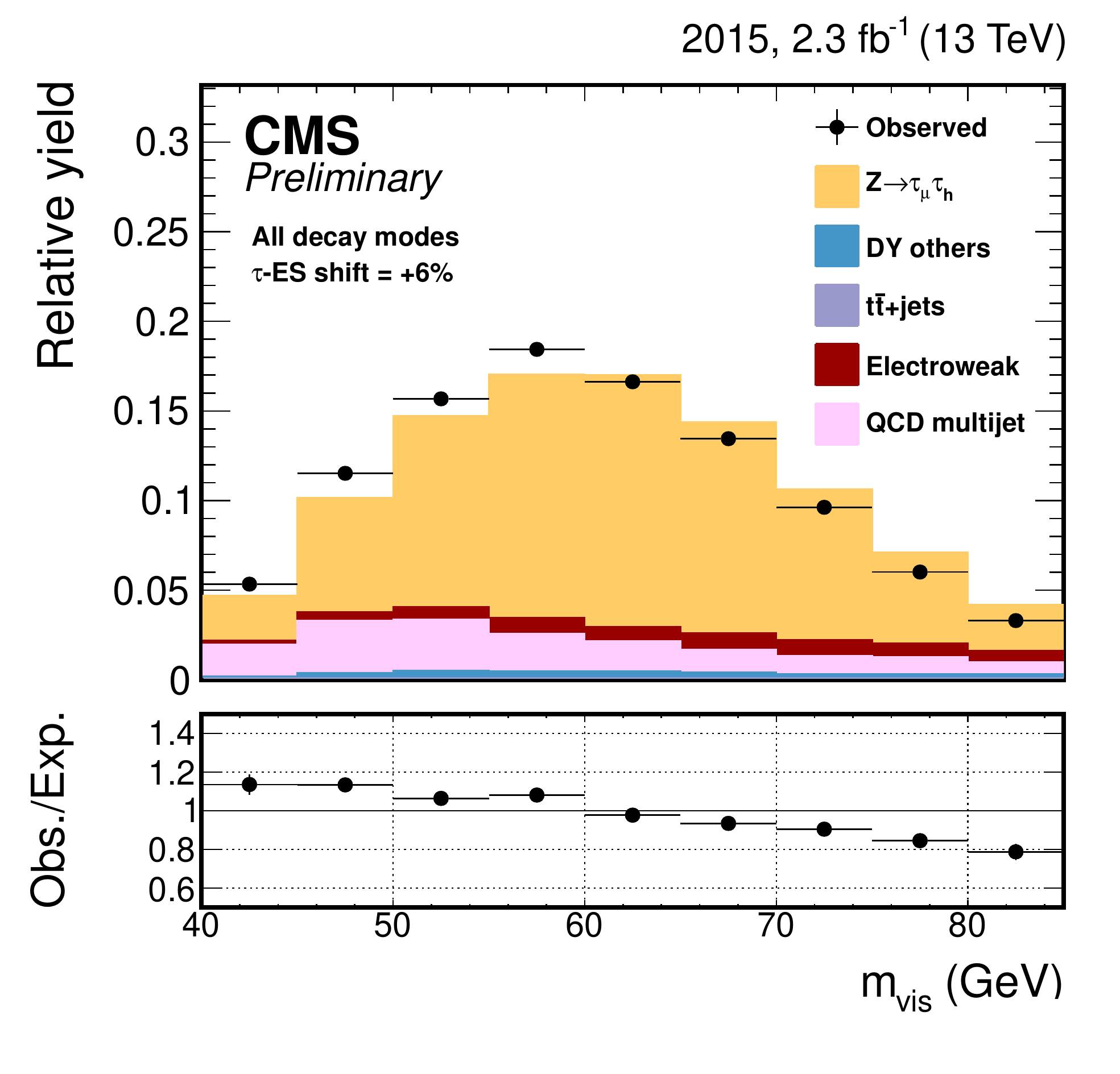}
\caption{CMS templates using simulation with $-6\%$, $0\%$ and $+6\%$ shift in tau energy scale compared to collision data after a $Z \to \tau\tau \to \mu \tauhad+3\nu$-enhancing selection~\cite{CMS:2016gvn}.}
\label{fig:cms_cal}
\end{figure}

As \tauhad\ are formed combining (calibrated) PF objects, no additional calibration procedure is required. The response is close to unity, see Figure~\ref{fig:atlas_cms_cal}.
A correction for small residual differences between data and
simulation is derived by fitting $m_{\tauhad}$ and $m_\mathrm{vis}$, the combined mass of the \tauhad\ and the muon candidate, separately for different decay modes~\cite{CMS:2016gvn}.
Templates with altered energy scales are applied following a $Z \to \tau\tau \to \mu \tauhad+3\nu$ selection to extract a best-fit value, see Figure~\ref{fig:cms_cal}. 
The derived corrections are below $1\%$.

\section{Tau performance measurements}
\begin{figure}[htb]
\centering
\includegraphics[width=0.33\textwidth]{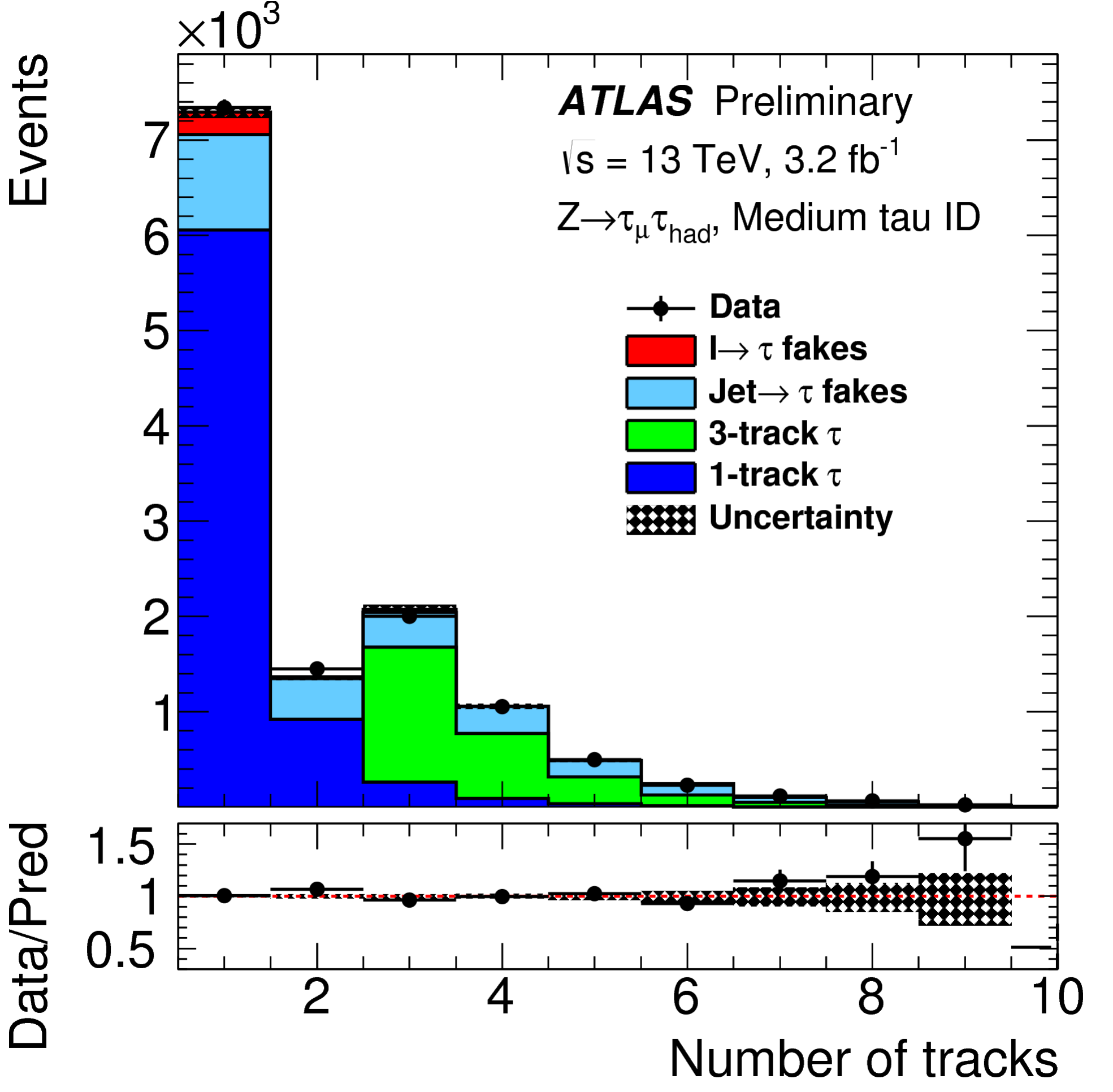} 
\includegraphics[width=0.35\textwidth]{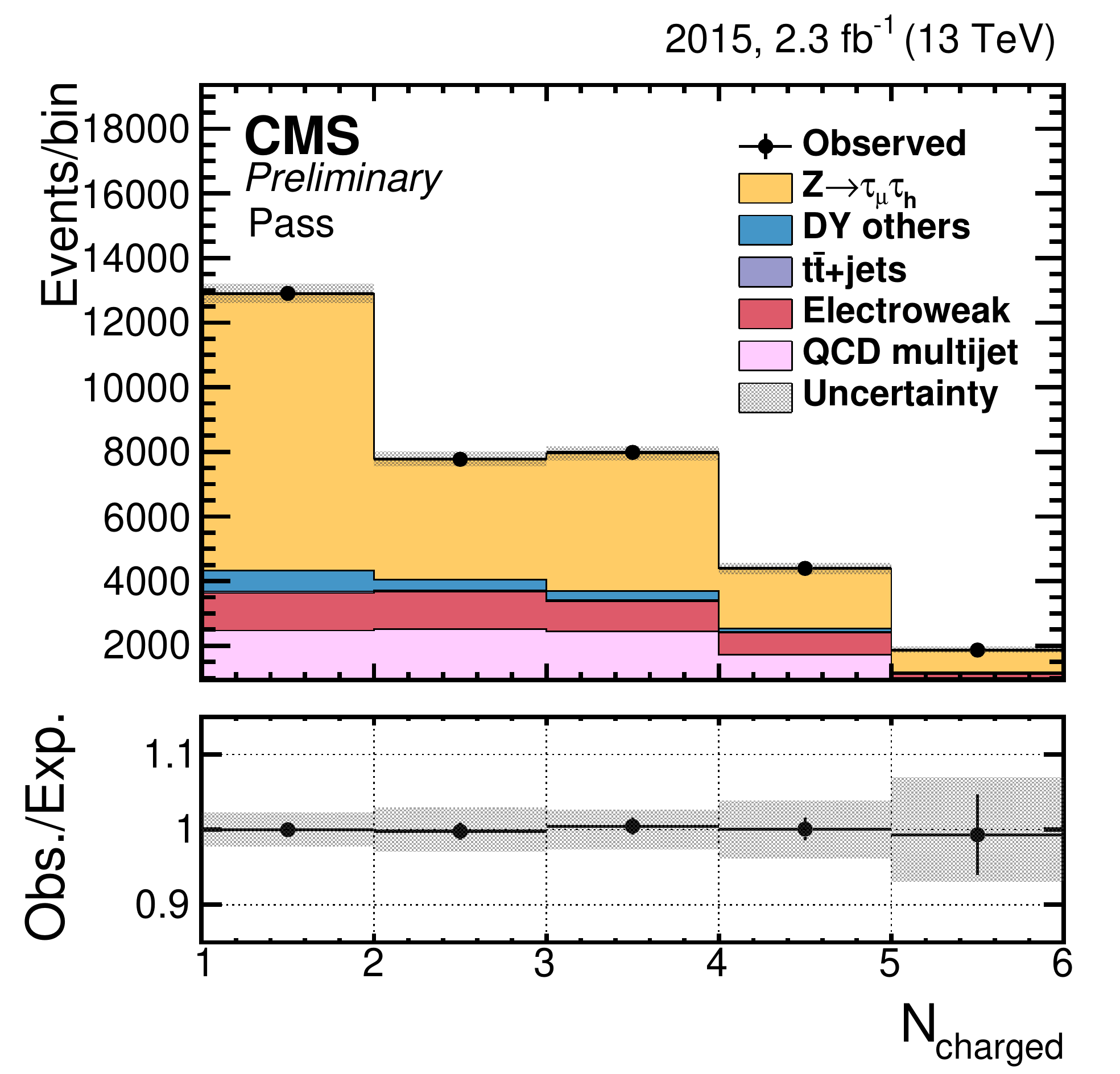} 
\caption{Number of tracks after passing $\tau$ identification criteria, ATLAS~\cite{ATLAS:2017mpa} (left) and CMS~\cite{CMS:2016gvn} (right).}
\label{fig:perf1}
\end{figure}

The standard candle for \tauhad\ measurements are $Z \to \tau\tau \to \mu\tauhad+3\nu$ events, using a tag-and-probe technique with a tag muon. 
%
%
The identification efficiency is measured with a combined fit e.g. to the number of charged hadrons (tracks) distribution before and after requiring identification criteria, 
see Figure~\ref{fig:perf1}. Both for ATLAS and CMS, the measured efficiencies agree with the expectation within uncertainties. Complementary studies are done in 
regions enhancing $W \to \tau\nu$ or $t\bar{t}$ events and by considering ratios of $Z \to \mu\mu$ and $Z \to \tau\tau$ events.

\begin{figure}[t]
\centering
\includegraphics[width=0.33\textwidth]{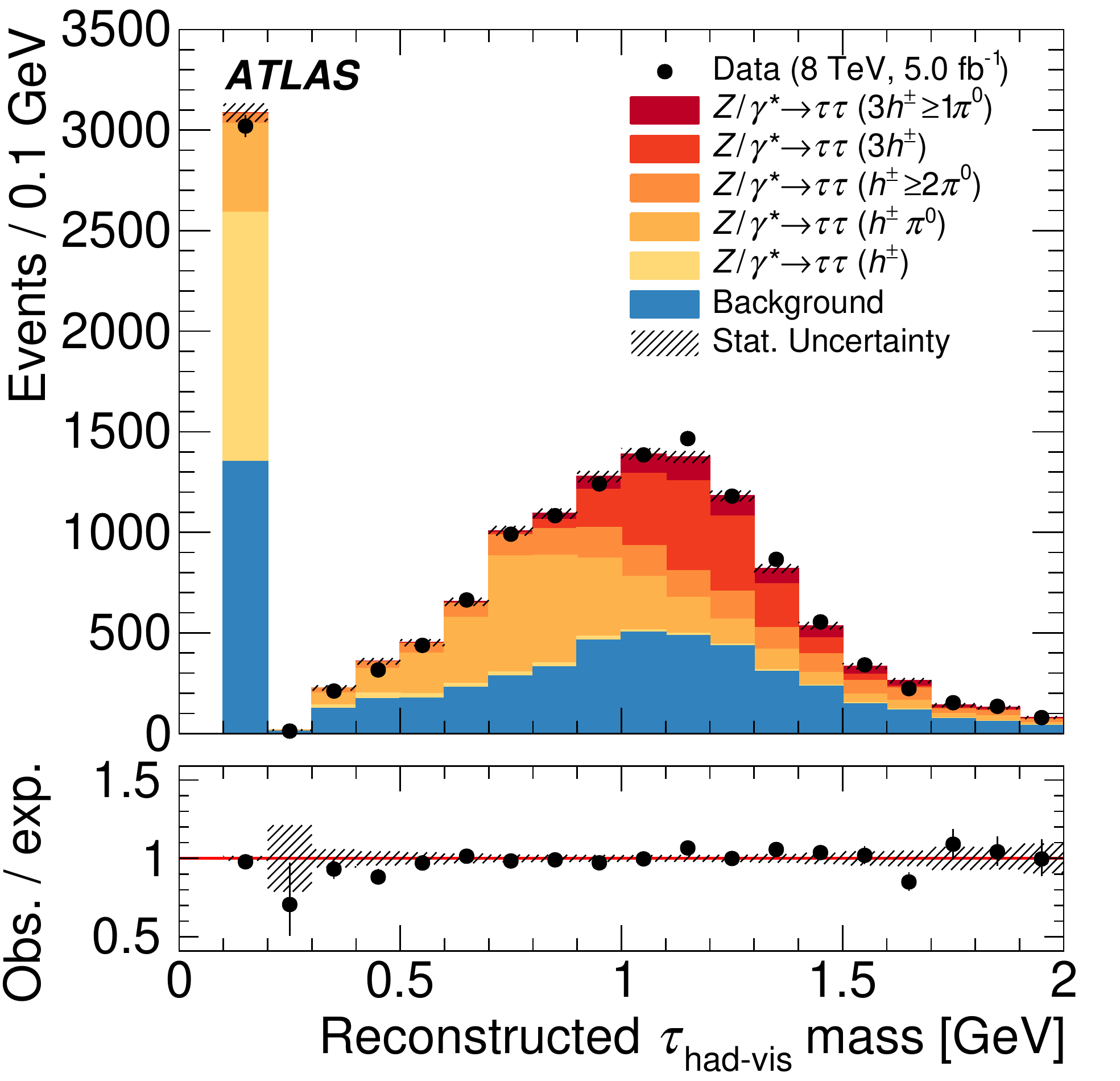}
\includegraphics[width=0.34\textwidth]{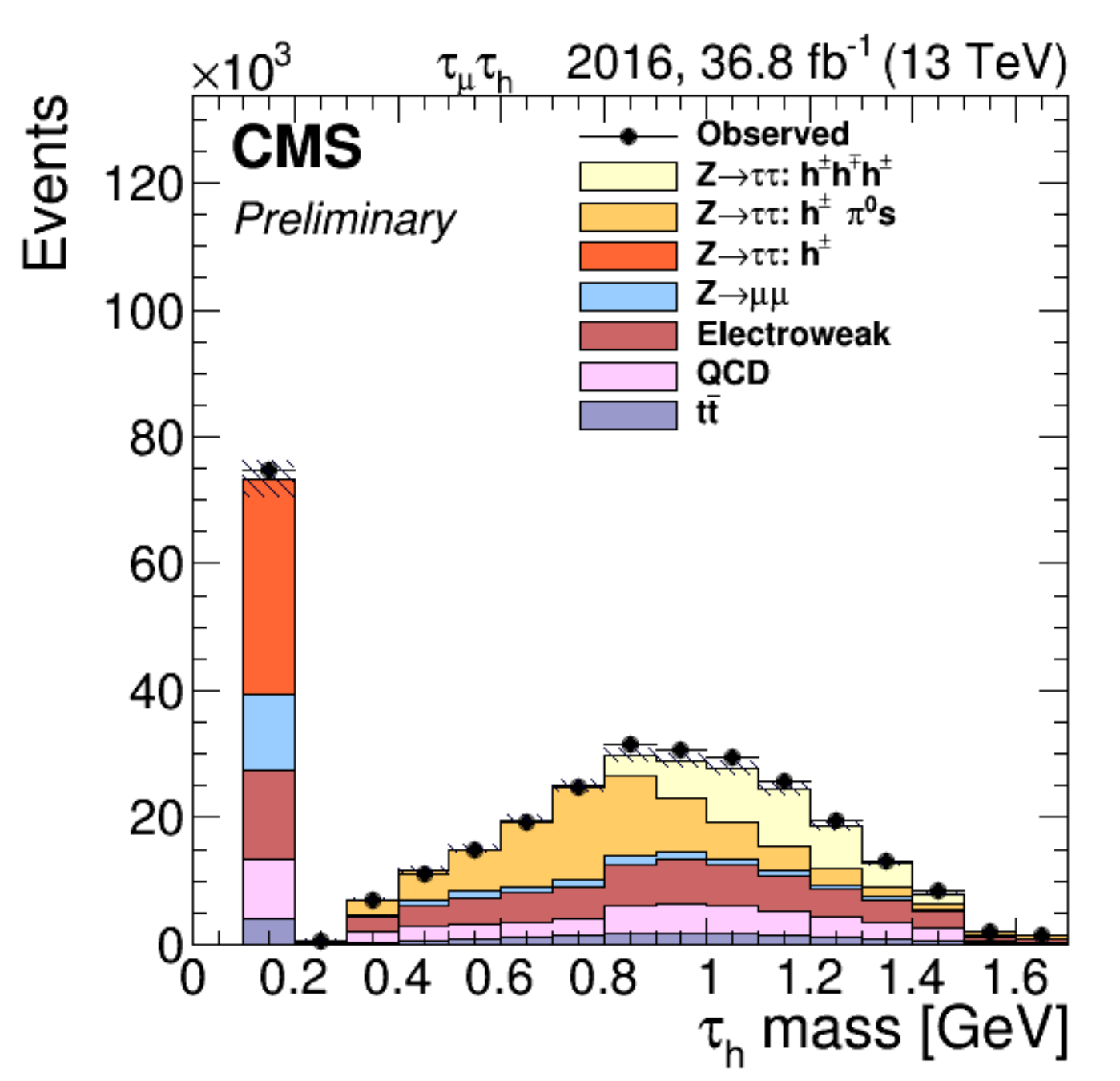}
\caption{\tauhad\ mass after passing $\tau$ identification criteria, ATLAS~\cite{Aad:2015unr} (left) and CMS~\cite{cms_mtau} (right).}
\label{fig:atlas_cms_taumass}
\end{figure}

The $\tau$ energy scale and resolution are most commonly tested by reconstructing either $m_{\tauhad}$ or $m_\mathrm{vis}$ following a 
selection to enhance $Z \to \tau\tau \to \mu \tauhad+3\nu$ events. The $\tauhad$ mass is shown in Figure~\ref{fig:atlas_cms_taumass} for 2012 data (ATLAS) 
and 2016 data (CMS). Both resolution and scale uncertainty are comparable for the two experiments. The resolution is roughly 10\%, 
depending strongly on e.g. $\pt$ and $\eta$. The scale uncertainty is about $1\%-3\%$, depending on decay mode, when accounting for differences 
of data and 
modeling. ATLAS also measures the absolute scale uncertainty to be between $2\%-6\%$, depending on $\pt$, $\eta$ and decay mode.



\section{Conclusions}
The identification of hadronic $\tau$ lepton decays is an integral part of the LHC program. ATLAS and CMS use different approaches 
but their methods have recently started converging (particle flow usage at ATLAS, and increased MVA usage at CMS). Overall, 
in spite of a hostile high-pile-up environment they both deliver excellent performance well beyond the expectations before the LHC start.

\end{document}

%% file: econfmacros.tex
\def\beq{\begin{equation}}
\def\eeq#1{\label{#1}\end{equation}}
\def\eeqn{\end{equation}}

\def\beqa{\begin{eqnarray}}
\def\eeqa#1{\label{#1}\end{eqnarray}}
\def\eeqan{\end{eqnarray}}

\let\bar=\overbar

\def\Dslash{\not{\hbox{\kern-4pt $D$}}}
\def\dslash{\not{\hbox{\kern-2pt $\del$}}}

\def\msb{{\bar{\ssstyle M \kern -1pt S}}}




%% file: eprint.bbl
\begin{thebibliography}{99}


\bibitem{atlas}
  ATLAS Collaboration,
  The ATLAS Experiment at the CERN Large Hadron Collider,
  JINST {\bf 3} (2008) S08003
\bibitem{cms}
  CMS Collaboration,
  The CMS experiment at the CERN LHC,
  JINST {\bf 3} (2008) S08004  

\bibitem{atlas_run2}
  ATLAS Collaboration,
  Reconstruction, Energy Calibration, and Identification of Hadronically Decaying Tau Leptons in the ATLAS Experiment for Run-2 of the LHC,
  ATL-PHYS-PUB-2015-045
  [http://cds.cern.ch/record/2064383].

\bibitem{ATLAS:2017mpa}
  ATLAS Collaboration,
  Measurement of the tau lepton reconstruction and identification performance in the ATLAS experiment using $pp$ collisions at $\sqrt{s}=13~{\rm TeV}$,
  ATLAS-CONF-2017-029
  [http://cds.cern.ch/record/2261772].

\bibitem{CMS:2016gvn}
  CMS Collaboration,
  Performance of reconstruction and identification of tau leptons in their decays to hadrons and tau neutrino in LHC Run-2,
  CMS-PAS-TAU-16-002
  [http://cds.cern.ch/record/2196972].

\bibitem{Aad:2015unr}
  ATLAS Collaboration,
  Reconstruction of hadronic decay products of tau leptons with the ATLAS experiment,
  Eur.\ Phys.\ J.\ C {\bf 76} (2016) 295
  [arXiv:1512.05955].

\bibitem{Khachatryan:2015dfa}
  CMS Collaboration,
  Reconstruction and identification of $\tau$ lepton decays to hadrons and $\nu_\tau$ at CMS,
  JINST {\bf 11} (2016) P01019
  [arXiv:1510.07488].



%
\bibitem{cms_mtau}
  CMS Collaboration,
  Tau-Id performance with full 2016 dataset using $Z\to\tau_\mu \tau_\mathrm{h}$ events,
  CMS-DP-2017-002
  [http://cds.cern.ch/record/2243476].


  



\end{thebibliography}
